\documentclass[useAMS,usenatbib,usegraphicx]{mn2e}

\newcommand{\kms}{~km s$^{-1}$~}
\newcommand{\dotM}{~M$_{\odot}$~yr$^{-1}$~}
\newcommand{\WR}{WR~147~}
\newcommand{\xspec}{{\small ~XSPEC~}}

\title[CSW models with NEI: X-rays from \WR ]
{Colliding Stellar Wind Models with 
Nonequilibrium Ionization: X-rays from \WR }
\author[S. A. Zhekov]
{Svetozar A. Zhekov\thanks{E-mail: szhekov@space.bas.bg}\\
Space Research Institute, Sofia-1000, Moskovska str. 6, Bulgaria}

\begin{document}

\date{}

\maketitle

\label{firstpage}

\begin{abstract}
The effects of nonequilibrium ionization are explicitly taken into 
account in a numerical model which describes colliding stellar winds
(CSW) in massive binary sytems.  This new model is used to analyze 
the most recent X-ray spectra of the WR$+$OB binary system \WR.
The basic result is that it can adequately reproduce the observed 
X-ray emission (spectral shape, observed flux) but some adjustment 
in the stellar wind parameters is required. Namely, (i) the stellar 
wind velocities must be higher by a factor of $1.4 - 1.6$;
(ii) the mass loss must be reduced by a factor of $\sim 2$.
The reduction factor for the mass loss is well within the
uncertainties for this parameter in massive stars, but given the fact 
that the orbital parameters (e.g., inclination angle and eccentricity) 
are not well constrained for \WR, even smaller corrections to the mass 
loss might be sufficient. Only CSW models with nonequilibrium ionization 
and equal (or nearly equal) electron and ion postshock temperature 
are successful.  Therefore, the analysis of the X-ray spectra of \WR
provides evidence that the CSW shocks in this object must be 
{\it collisionless}.
\end{abstract}


\section{Introduction}
\label{sec:intro}
The Wolf-Rayet (WR) star \WR (AS 431) is a classical example of colliding 
stellar wind (CSW) binary: it possesses relatively strong X-ray
emission and it is a nonthermal radio source. 
Probably the most interesting piece of information about this WR$+$OB
binary system comes from the radio, namely, high-resolution
interferometer observations have spatially resolved its
emission into two components: a southern thermal source, WR~147S
(the WN~8 component in the binary), and a
northern nonthermal source, WR~147N 
(\citealt{ab_86}; \citealt{mo_89}; \citealt{ch_92};
\citealt{con_96}; \citealt{wi_97}; \citealt{sk_99}, hereafter S99).
Moreover, long-term radio variability has been reported both from 
the thermal \citep{con_99} and nonthermal \citep{se_01}
sources. The radio spectrum of WR~147S has a spectral index 
$\alpha = +0.6; +0.62$~($S_{\nu} \propto \nu^{\alpha}$),
which is very close to the canonical value for the thermal free-free
emission from ionized stellar winds (\citealt{ch_92}; \citealt{wi_97};
S99). On the other hand, the nonthermal
radio emission of WR~147N cannot be described by a simple power law,
which is expected from synchrotron models that assume power-law
electron energy distribution (S99), and more
elaborated modelling is required (\citealt{do_03}; \citealt{pitt_06}). 

High-resolution infrared images clearly showed the NIR counterparts of
the two radio sources located at $\approx 0\farcs64$ from each
other and WR~147N was classified as a B0.5V star \citep{wi_97}
but an earlier spectral class, O8-O9 V-III, was suggested 
from the {\it Hubble Space Telscope} observations \citep{nie_98}.
Given the distance to \WR of $630\pm70$~pc \citep{ch_92},
the projected (or minimum) binary separation is $403\pm13$~au.
It is worth noting that analyzing radio and IR images,
\citet{wi_97} have found that the WR~147N radio peak is
displaced by $0\farcs07$ southward from its IR counterpart.
Although this value is within the uncertainties of the data, if it is
real, then the nonthermal radio source can be associated with the CSW
region located off the surface of the OB companion. This physical picture
found further support from the high spatial resolution {\it Chandra} 
image, showing that the X-ray emission peak lies at or near
WR~147N \citep{pitt_02}.

The X-ray observations of \WR have revealed the presence of thermal
emission from high temperature plasma but they also demonstrated 
how it is imporatnt to have data with good quality. 
Namely, \WR was first detected by the
{\it Eistein} observatory and it was concluded that the X-rays are due
to plasma at temperature $kT \geq 0.5$~keV \citep{cai_85}. 
Analysis of the moderate
resolution CCD spectra, obtained with {\it ASCA} and having not very
high photon statistics, suggested that the X-rays likely originate in
multitemperature plasma whose cooler component has a temperature of
$kT \approx 1$~keV, but the characteristics of the hotter component
could not be constrained (S99). Finally, 
the higher signal-to-noise {\it XMM-Newton} data of \WR 
presented by Skinner et al. (2007, hereafter S07) 
have undoubtedly revealed the elusive high temperature plasma in this
object by detecting the Fe K$\alpha$ complex at 6.67~keV. The
corresponding analysis of the X-ray spectra has shown that the plasma
tempeature is as high as $kT = 2.7$~keV. 
The authors pointed out that neither long-term X-ray variability 
nor changes in the X-ray absorption were detected. Thus, the 
`evolution' of the plasma characteristics, mentioned above, is merely 
a result of the acquisition of higher quality data with each successive
generation of X-ray satellites.

Apart from the clear evidence for presence of high temperature plasma
in \WR, S07 have found indications that 
further improvement of numerical CSW models is needed. Namely, 
higher terminal wind velocities are required to account for the hot
plasma detected in this object, and future CSW models must include 
some additional physics such as nonequilibrium ionization (NEI).

Motivated by this, the aim of this study was to develop new CSW models 
that take into account the NEI effects and then to confront them with 
the high quality {\it XMM-Newton} spectra of \WR. The model details 
are given in \S~\ref{sec:model}. The results from the model application 
are presented in \S~\ref{sec:res} and discussed in \S~\ref{sec:disc}.
Our conclusions are found in \S~\ref{sec:con}.

\section{Data}
\label{sec:data}
The X-ray observations  of \WR were performed with the {\it
XMM-Newton} observatory in November 2005. All the details about
the EPIC data and the data reduction are found in S07.
We only note that our analysis is based on the
PN and MOS spectra  of \WR and the latter is the sum of the
spectra from the two MOS detectors.
Thus, each analyzed spectrum has about the same number of counts
(4436 in PN; 4148 in combined MOS). They were
rebinned to have a minimum of 20 cts per bin.
All spectral fits were performed in the recent version (11.3.2)
of the software package for analysis of X-ray spectra, \xspec
\citep{a_96}.  The spectra
were analyzed simultaneously with exactly
the same model and only the \xspec
normalization parameter was decoupled to allow for any
calibration difference between the PN and MOS detectors.

\section{Model}
\label{sec:model}
The goal of this study is to use a realistic model that determines
the physical parameters of the X-ray emitting plasma, and on its basis
to simulate the X-ray spectra to be confronted with the
observational data.
A good starting point is to consider a simple model which describes
the basic physical situation and then
elaborate on those that include more complex physics (e.g., nonuniform
gas flows, thermal conduction etc.) which might be potentially important 
for the case under consideration. 
As in the case of other wide stellar binaries (e.g., WR~140; 
\citealt{zhsk_00}), an adiabatic model of CSW interaction was adopted
for describing this phenomenon in \WR while more complicated models are
beyond the scope of the present work.
Adiabatic CSWs also
have the advantage that the hydrodynamic solution in the interaction
region does not depend on such details as: e.g., exact metal
abundances, whether or not the plasma is in ionization equilibrium, 
or any difference between electron and ion temperatures. On the other hand,
any of these may directly affect the shape of the emitted X-ray spectrum
and the strength of various emission lines. Thus, it might be
necessary to take them into account when simulating the X-ray emission 
from such objects. We note that all these physical
characteristics allow the X-ray spectral simulations to be decoupled
from the hydrodynamic modelling of CSWs in wide binaries. Namely,
the temperature and density distributions of the hot plasma, as
initally derived from the numerical hydrodynamics, can be further 
postprocessed to model the corresponding X-ray emission.
We next describe the steps in such an approach as applied to 
the case of \WR .

\subsection{CSW Model}
\label{subsec:csw_model}
A primary step is to determine the distribution of temperature and density
of the X-ray emitting plasma. For this purpose,
the hydrodynamic model of adiabatic CSWs by 
Myasnikov \& Zhekov (1993, hereafter MZh93) was used to
derive the physical parameters in the interaction region of \WR.
This model assumes spherical symmetry of the supersonic stellar winds
that have attained their terminal velocities in front of the shocks.
This assumption is well justified for wide stellar binaries
where neither the radiative braking nor the orbital motion is expected
to play an important role for the wind dynamics.
In this case, the interaction region has cylindrical symmetry and the
gasdymanic problem is determined entirely by the stellar wind
parameters and the binary separation. We note that the `shock
fitting' technique is used in this numerical model, and thanks to this 
an exact solution to all discontinuity surfaces (the two shocks and the
contact discontinuity) is derived. This means that there is no
numerical `mixing' between the shocked gases of the stellar winds
which further facilitates modelling the X-ray emission from the hot
plasma, especially, when the different chemical composition is
concerned.
The procedure of how the
effects of different electron and ion temperaures can be taken into
account is described in \citet{zhsk_00} and that for
handling the nonequilibrium ionization is given below.

\subsection{NEI Effects}
\label{subsec:nei}
In many astrophysical examples, the plasma emission is
determined by the state of ionization equilibrium. If plasma
characteristics such as temperature and density evolve on 
relatively small timescales, some time is also needed for a new
state of ionization equilibrium to be established. 
Therefore, we might have a chance to witness transition phases in the
plasma evolution. One such case is when the astrophysical plasmas 
are heated by shocks: gas temperature and density experience abrupt 
change. Fast shocks, as in supernova remnants or CSW binaries, can result
in high plasma temperatures, and the corresponding
hot-plasma emission is entirely determined by the collisional
processes of ionization and excitation mostly with electrons.
A detailed description of related processes that determine the
hot-plasma emission both in and out of ionization equilibrium can
be found in the review papers by \citet{li_99} and \citet{me_99}.

Obviously, the transition phases, such as the state of nonequilibrium
ionization, are important in evolving systems. On the other
hand, the NEI effects might play an important role also in
steady state conditions as in the CSW shocks in wide binaries.
Namely, despite the fact that the general picture does not change
in time, each individual parcel of gas follows the same `evolutionary'
pattern: it gets heated on the shock front, and then its ionization
state relaxes downstream towards the corresponding equilibrium
conditions: collisional ionization equilibrium (CIE). Whether or not 
the transition phase is important depends on how quickly CIE is
established with respect to the typical timescale of the flow.

In order to determine the relative importance of the NEI effects
in CSW binaries,
we consider a dimensionless parameter $\Gamma_{nei} =
\tau/\tau_{nei}$. Here $\tau$ is the timescale of gasdynamics and
$\tau_{nei}$ is the characteristic time to establish ionization
equilibrium, which we can refer to as the NEI timescale. 
The NEI effects must be taken into account if $\Gamma_{nei} \leq 1$
but can be neglected for $\Gamma_{nei} \gg 1$.
In the hydrodynamic model at hand, $\tau = D/V_{\infty}$,
where $2D = a$ is the separation between the binary components and 
$V_{\infty}$ is the terminal stellar wind velocity. Unfortunately, 
there is no `unique' NEI timescale and such a parameter can be 
introduced for each
chemical element whose ionization state is considered. Although the
latter requires one to solve a set of differential equations,
qualitative considerations show that CIE is controlled mostly by the
rate coefficients of collisional ionization and recombination to
hydrogen- and helium-like species of a given chemical element.
This is particulary the case for hot, X-ray emitting plasmas that are
of interest in this study. Guided by this (see also the NEI discussion 
and the case example of oxygen in \citealt{li_99}), we can introduce a 
representative NEI timescale $\tau_{nei} = 10^{12}/n_e$, where $n_e$
is the electron number density of the hot plasma. In that case
$\Gamma_{nei}$ can be expressed in terms of the basic CSW parameters:
\begin{equation}
  \Gamma_{nei} = 
           1.21 \frac{\chi_e \dot{M}_5}{\bar{\mu} V^2_{1000} D_{16}}
\label{eq:gam_nei}
\end{equation}
where $\dot{M}_5$  is the mass-loss rate in units of 10$^{-5}$ \dotM,
$V_{1000}$ is the stellar wind velocity in units of 1000\kms, 
$D_{16} = D/10^{16}$~cm, $\bar{\mu}$ is the mean atomic weight for
nucleons, and $\chi_e = n_e/n $ is the relative electron number
density ($n$ is the nucleon number density).

Given the stellar wind and binary parameters of \WR 
(see \S~\ref{subsec:wind}),
it is anticipated that the NEI dimensionless parameter 
$\Gamma_{nei} \approx 1$ and $\Gamma_{nei} \ll 1$
for the shocked WR and O wind, respectively. This means that the NEI
effects likely play an important role for the X-ray emission
from the CSW region in this object.

Keeping in mind the cylindrical symmetry of the hydrodynamic problem,
a full set of 2D, time-dependent partial differential equations must be
considered if the NEI effects need be taken into account in the 
CSW models.
From a numercial view point, an alternative to this is to
consider the ionization balance along the flow streamlines. In this
case the numerical task is much simpler and a set of ordinary
differential equations represents the evolution of the ionization state
of each chemical element:
\begin{equation}
  \frac{d n_i}{d\tau_e} = C_{i-1} n_{i-1} - (C_i + \alpha_i) n_i +
                          \alpha_{i+1} n_{i+1}
\label{eq:nei}
\end{equation}
where $C$ and $\alpha$ denote the collisional ionization and
recombination rate coefficients, respectively, and 
$n_{i-1,i,i+1}$ are the number densities of any three consecutive 
ionization stages of a given chemical
element. The parameter $\tau_e$ is the so-called ionization 
time, $\tau_e = \int{n_e dt}$, and the integration 
starts at the shock
front and is carried out downstream along a specific streamline.

It is well known that the ionization equations (eqs.~[\ref{eq:nei}]) 
are a set of {\it stiff} ordinary differential equations, and various 
techniques have been developed to deal with this numerical problem. 
Very often an
eigenfunction method (e.g. \citealt{hu_85}) is used to solve
these equations efficiently, and this method was adopted in various
NEI models available in \xspec (\citealt{bo_01}).
Therefore, the latter was the basis for incorporating our NEI CSW 
models in the spectral fits to the X-ray data of \WR.
We note that the NEI models in \xspec usually assume
constant electron density and temperature while this is not the case
in the density and temperature stratified interaction region in CSW
binaries. To overcome this problem, we adopted a piecewise integration
approach. Namely, the hydrodynamic model allows for deriving all the
necessary plasma parameters (e.g., number density, electron temperature,
ionization time) at the grid points along a given streamline of the
flow. It is thus assumed that density and temperature are constant
in between any two consecutive ($j$ and $j+1$) grid points along the 
streamline and now the NEI problem can be solved 
using the corresponding 
models available in \xspec
(with the only modification that at each step the initial values for 
eqs.~[\ref{eq:nei}] are those from the previous grid point).
Once this is done, the X-ray spectrum of
the hot plasma along a given streamline is derived and the sum over all
the streamlines gives the total X-ray emission from the interaction
region.

\subsection{CSW Models in \xspec}
\label{subsec:xspec}
In order to make a direct comparison between the observations and 
the X-ray emission predicted by the CSW models, we developed a new
spectral model for \xspec. The X-ray emission from the interaction 
region is thermal, thus, its characteristics are determined by
the temperature, emission measure and chemical abundances  of the hot
plasma. This is true for the case of CIE and an additional parameter
(ionization time) should be provided as well, if the NEI effects are 
important. All these necessary ingredients are given by the
hydrodynamic CSW model. For modelling the X-ray emission in CIE
the \xspec optically-thin plasma model {\it apec} is used at each
grid point of the interaction region,
while the case of NEI is considered as described above 
(\S~\ref{subsec:nei}). Such an approach also has the advantage
that the fits with theoretical spectra of CSW binaries can 
be easily confronted with the results from the standard
fits that usually make use of discrete-temperature plasma models 
(both in and out of ionization equilibrium) available in \xspec
(e.g., {\it vmekal, vapec, vpshock}).
This, in turn, will be helpful for interpreting the X-ray data
from CSW binaries when detailed numerical modelling is not feasible 
for various reasons (e.g., data quality, binary parameters are not 
available).

Two types of test were performed for the new spectral models. First,
for internal consistnecy, the distribution of emission measure of the
hot plasma, as derived on the original grid of the hydrodynamic CSW
model, was compared with the one resulting from the intergration along
the flow streamlines. Second, results from our
piecewise integration approach were confronted with the ones from the
\xspec {\it vpshock} model. These tests made us confident about the
further use of the spectral CSW models for analyzing the {\it
XMM-Newton} data of \WR.

Finally, using the known distance to the observed  object, 
the emission measure from the hydrodynamic CSW model can be 
converted into \xspec units
\footnote{
Conversion of emission measure into \xspec units
is done by multiplying it with the term $10^{-14}/4 \pi d^2$, where $d$ 
is the distance to the object in cm. This term comes from the \xspec
normalization factor of the flux for optically-thin plasma models.
}.  
In such a case, the \xspec {\it norm} 
parameter of our spectral model would have values close to unity 
if theoretical and observed fluxes match each other. Deviations from 
unity may be an indication of inaccuraices in the adopted stellar wind
parameters or the distance to the object.

\subsection{Basic Parameters for the CSW Model in \WR}
\label{subsec:wind}
For consistentcy with previous works (S99, S07), we define a
`standard' CSW model for \WR, which will also serve as a starting
point for the fits to the X-ray emission from this object.
As already mentioned, the distance of 630 pc to \WR is adopted in
this study, and the
stellar wind parameters are: $\dot{M}(WR) = 4\times10^{-5}$\dotM,
$v_{\infty}(WR) = 950$\kms, $\dot{M}(O) = 6.6\times10^{-6}$\dotM,
$v_{\infty}(O) = 1600$\kms. These values define a wind momentum ratio 
$\eta = [\dot{M}(O) v_{\infty}(O)]/[\dot{M}(WR) v_{\infty}(WR)] =
0.028$~(note that $\Lambda = 1/\eta$ is used in MZh93 for the wind
momentum ratio, but for consistency with the work of other authors 
$\eta$ will be used throughout this study). 
Also, a minimum value of the stellar separation in this binary
system is $a = 2D = 403$~au, 
and we note that this parameter depends on the inclination
angle which is uncertain. Based on analysis of VLA radio images,
values of $i = 45\degr\pm15\degr$ \citep{con_99}
and $i \approx 30$\degr \citep{con_04} have been reported but 
other values may well be possible. It is so since the inclination angle
was derived by fitting a thin shell model to the radio image and it should
be kept in mind that such a model can not be considered satisfactory
for wide stellar binaries, where the CSWs are adiabatic, since it gives 
only the shape of the contact discontinuity surface. On the other hand,
the shape of the shock surfaces (e.g., their `opening angle') is quite
different which makes results from such a model application
higly uncertain. Therefore, 
values for the binary separation in \WR a few times larger than 
the minimum value given above are not ruled out.

Thus, we have all the necessary input 
parameters for the hydrodynamic CSW model at hand (MZh93), and
elemental abundances have to be provided for modelling the thermal
X-ray emission from the interaction region.
Due to their advanced evolutionary status, WR stars are objects with 
nonsolar abundances
and for this study the following set of them is adopted. The
abundances for H, He, Ne and Ca are based on the analysis of the IR
spectrum of \WR  \citep{mo_00},
and the other elements have values `typical'
for a WN star \citep{vdh_86}. For convenience, all values
are with respect to the solar number abundances (unity means solar 
abundance as in \citealt{an_89}):
H $= 1$, He $= 25.6$, C $= 0.9$, N $= 140$, O $= 0.9$, Ne $= 44.3$,
Mg $= 14.4$, Si $= 15.2$, S $= 3.6$, Ar $= 5$, Ca $= 4.7$, Fe $= 7$,
Ni $= 1$
~(no information is available for Ni and it was assumed solar but
we note that the spectral fits are not sensitive to its value).
Due to the high X-ray absorption, the portion of the spectrum with
good statistics ($E \geq 0.9$~keV) is sensitive only to the elemental 
abundances of Ne, Mg, Si, S, Ar, Ca, Fe, which are allowed to
vary in our fits.  In all fits the elemental abundances of the O wind
were strictly solar and they are kept fixed in the modelling of the X-ray
emission from \WR. 
The assumed WN abundances correspond to a relative electron density
$\chi_e = 1.71$, mean particle weight (electrons and nucleons) 
$\mu = 1.16$, and mean atomic weight for nucleons $\bar{\mu} = 3.14$.
For the O star the adopted abundances give $\chi_e = 1.1$, 
$\mu = 0.62$ and $\bar{\mu} = 1.3$.

\section{Results}
\label{sec:res}
The results from previous application of
the hydrodynamic CSW model for calculating the X-ray emission from \WR
indicated that there is a luminosity mismatch between the model
predictions and the observations (S99, S07). Namely, the theoretical
luminosity (or the amount, emission measure, of the hot plasma) is 
$\sim 3-4$ times larger than observed. Our new spectral
analysis confirmed that this difference between theory and observations 
is now valid for both cases of CIE and NEI, if the {\it nominal} 
wind parameters are used (e.g., the ones defined in 
\S~\ref{subsec:wind}). This conclusion also holds if the mass loss of
the O star, $\dot{M}(O)$, is varied. Two such cases were considered 
which have $\eta = 0.02$~($\dot{M}(O) = 4.7\times10^{-6}$\dotM)
and~$0.04$~($\dot{M}(O) = 9.4\times10^{-6}$\dotM), respectively 
(note that all other
parameters, $\dot{M}(WR)$, $v_{\infty}(WR)$, $v_{\infty}(O)$, kept
their {\it nominal} values, \S~\ref{subsec:wind}). Moreover, in all
the three cases, $\eta = 0.02, 0.028$~and~$0.04$, the overall quality
of the fits is not acceptable as indicated by the values of reduced 
$\chi^2_{\nu} = 1.63 - 1.93$ (with degrees of freedom $\nu = 332$ for
all the fits considered throughout this study). Therefore, it is
conclusive that the CSW model not only predicts an amount of X-rays 
larger than that observed for \WR, but it also does not give a good 
match to the exact shape of the observed X-ray spectrum, especially, 
in its high-energy part above 4 keV, where the theoretical spectrum is
softer than the observed one. 
Obviously, the data require presence of a hotter
plasma in the interaction region and, as already suggested, a remedy 
to this problem could be the assumption of higher values for the
stellar wind velocities (S07).
Note that the contribution of the shocked O wind is no more than
12-15\% at energies above 4 keV (S07 and this work), therefore, 
the velocity of the WN wind is a crucial parameter.

\begin{figure*}
\centering\includegraphics[width=3.0in, height=2.5in]{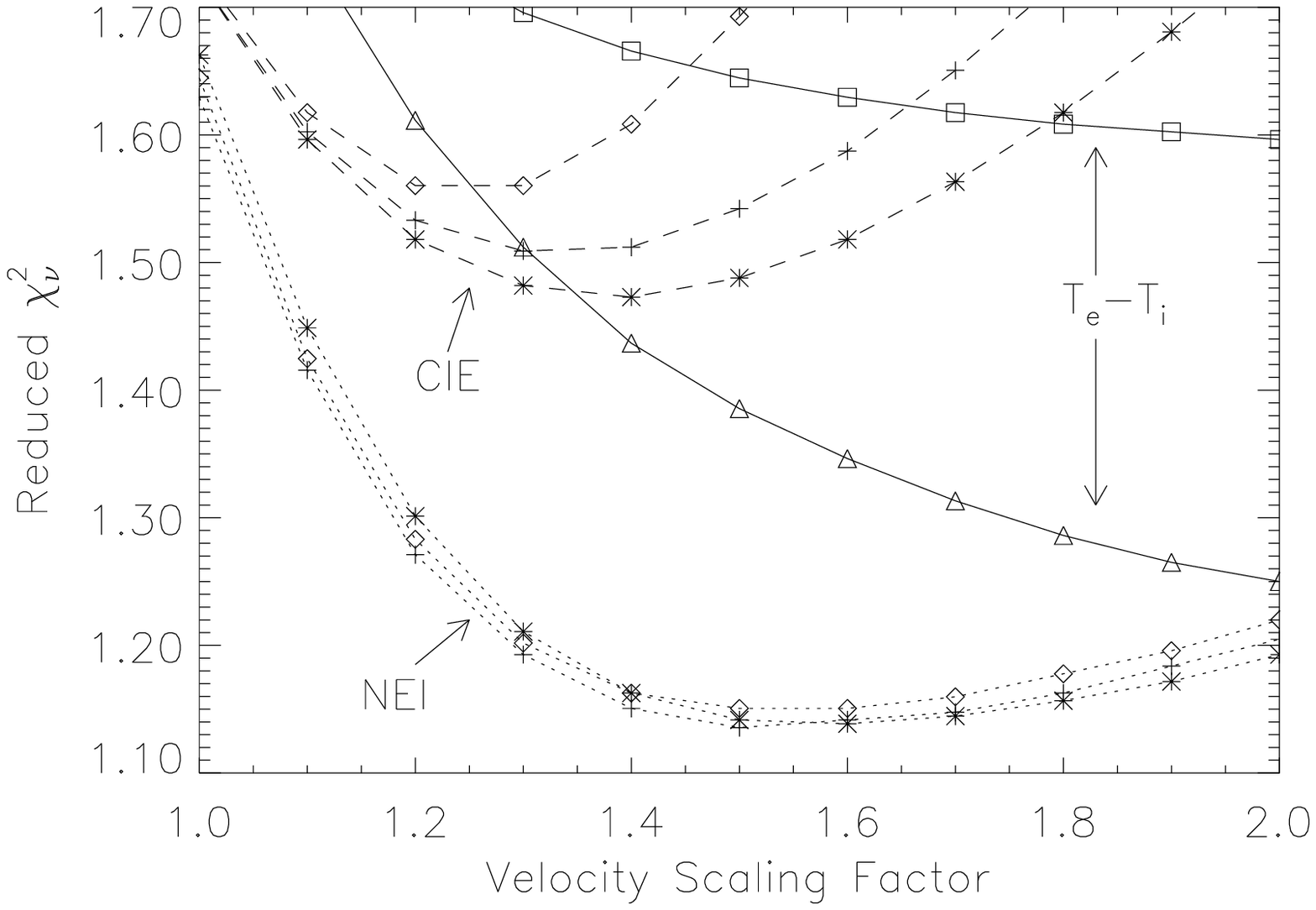}
\centering\includegraphics[width=3.0in, height=2.5in]{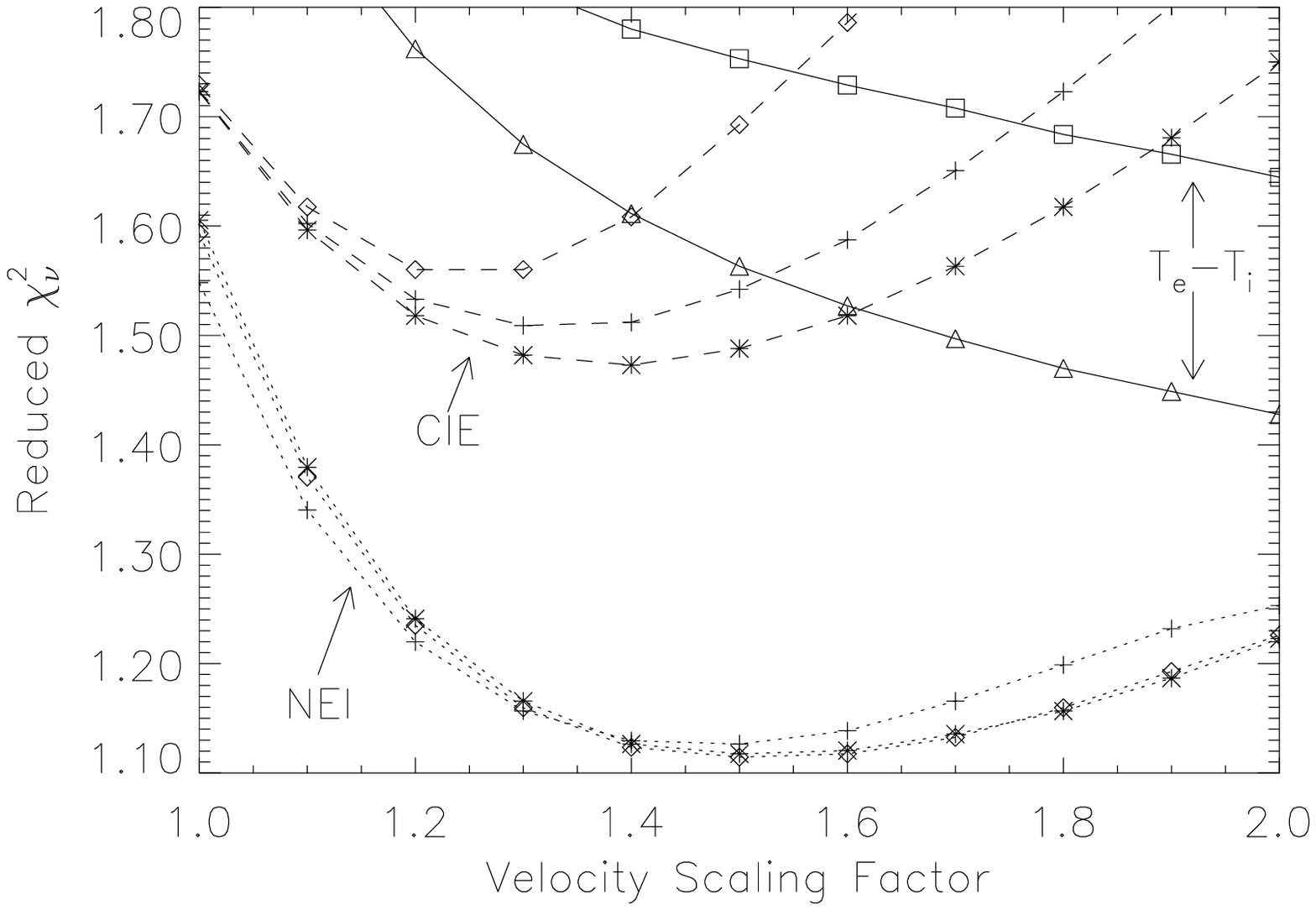}
\centering\includegraphics[width=3.0in, height=2.5in]{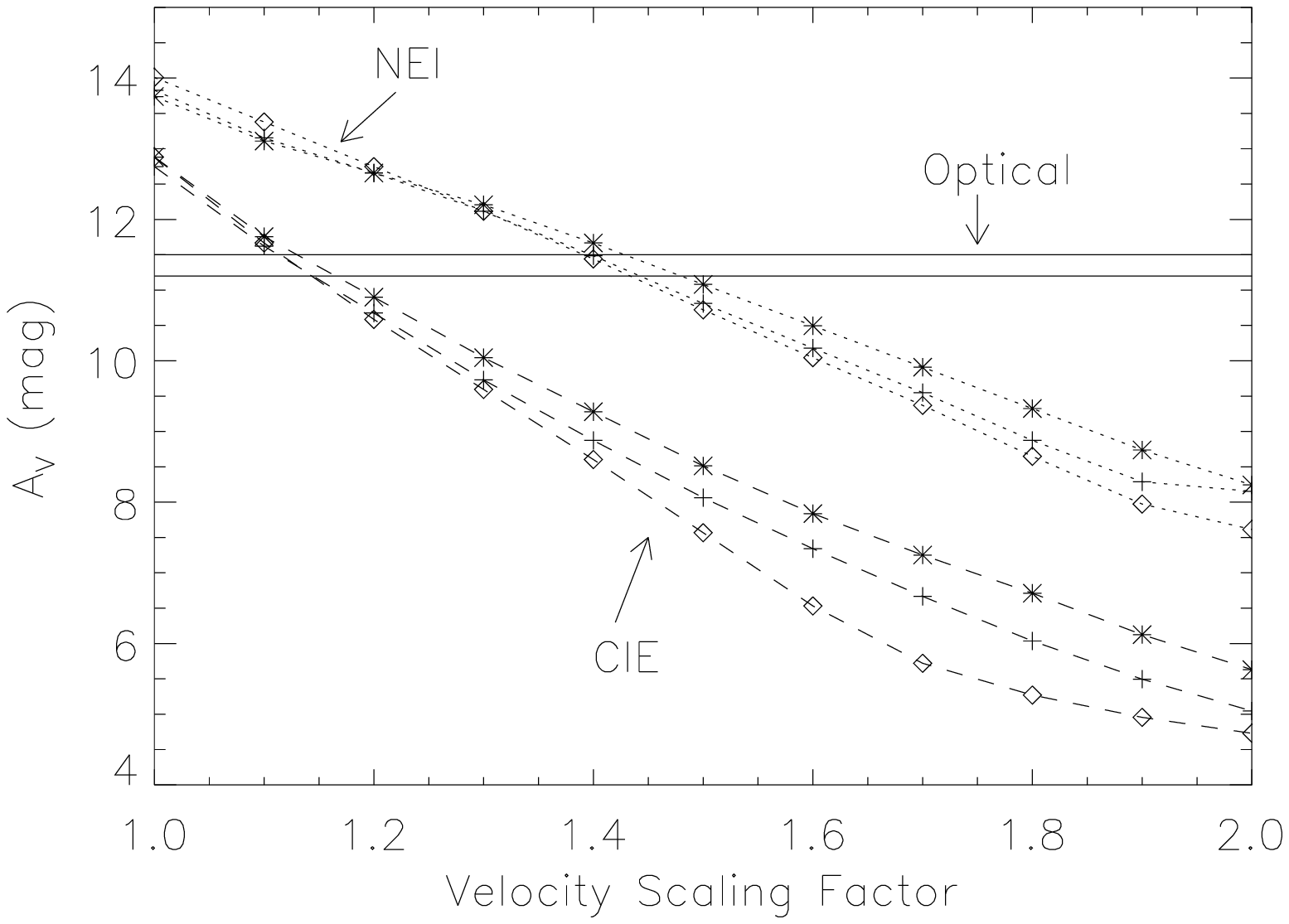}
\centering\includegraphics[width=3.0in, height=2.5in]{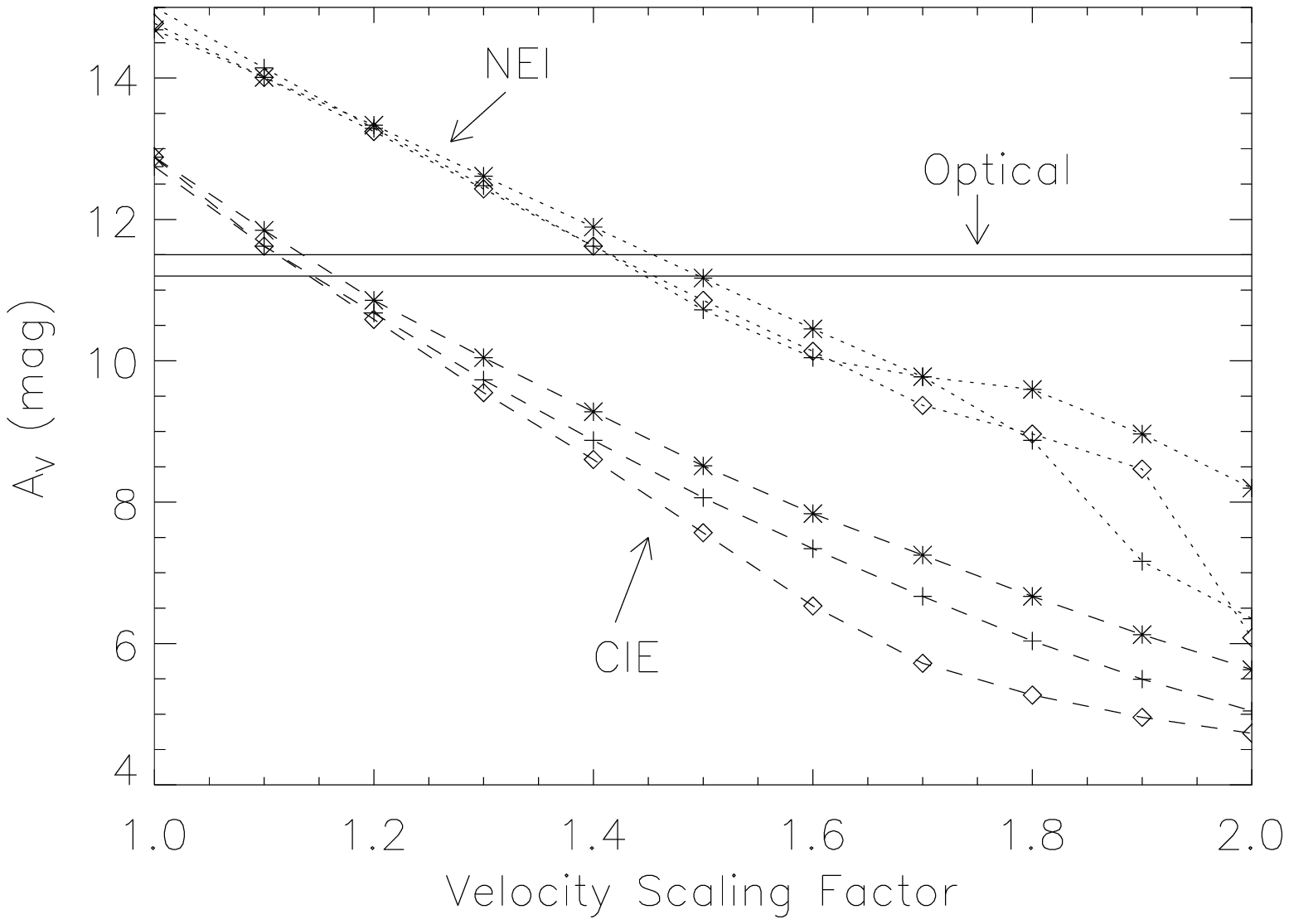}
\caption{Results from the CSW model fits. {\bf Left column:} Mass-loss 
rates scaled while the binary separation is kept at its `nominal' value. 
{\bf Right column:} The binary separation is scaled while mass-loss rates
have their `standard' values.
{\bf Upper panels:} Reduced $\chi^2_{\nu}$ vs. scaling factor for the
wind velocities. Dotted lines denote the models with nonequilibrium
ionization (NEI), and dashed lines present models that assume
the hot plasma is in collisonal ionization equilibrium (CIE). 
The plus signs are for the case of $\eta = 0.04$; asterisks mark models 
with $\eta = 0.028$; and diamonds present models with $\eta = 0.02$.
The solid lines illustrate the effect of nonequal electron and ion
temperatures ($\beta = T_e/T = 0.2$) for $\eta = 0.028$ as triangles
are for models with NEI while squares present those in CIE.
{\bf Lower panels:} The visual extinction as derived from the fits
with
various models where the $N_H$ to $A_V$ conversion formula from
Gorenstein (1975) was used. All symbols and lines have the same
meaning.
The two horizontal solid lines give the optical extinction deduced from
IR studies ($A_V = 11.5$ mag, Churchwell et al. 1992; $A_V = 11.2$
mag, Morris et al. 2000).
The degrees of freedom in all the fits are $\nu = 332$.
}
\label{fig:csw}
\end{figure*}
\begin{figure*}
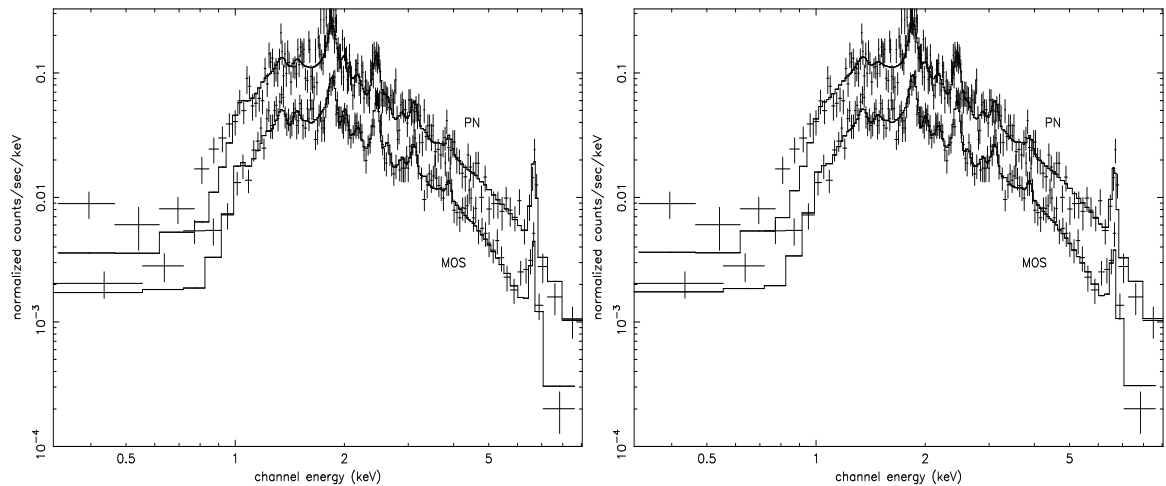

\centering\includegraphics[width=2.5in, height=3.0in,angle=-90]{fig2a.eps}
\centering\includegraphics[width=2.5in, height=3.0in,angle=-90]{fig2b.eps}
\caption{Examples of 
background-subtracted PN and MOS (MOS1$+$MOS2) spectra of \WR and
the corresponding best fit NEI CSW models for $\eta = 0.028$ and stellar
wind velocities scaled by a factor of 1.5. 
{\bf Left panel:} the case of reduced mass losses by a factor of 0.5. 
{\bf Right panel:} the case with increased binary separantion by a
factor of 4. 
Fit results are given in Table~\ref{tab:csw}.
}
\label{fig:spec}
\end{figure*}
\begin{table}
\begin{center}
\caption{Examples of CSW ($\eta = 0.028$) spectral fits for \WR }
\label{tab:csw}
\begin{tabular}{lll}
\hline
\hline
\multicolumn{1}{l}{} & \multicolumn{1}{c}{Mass loss$^{(a)}$}  &
\multicolumn{1}{c}{Binary separation$^{(b)}$} 
  \\
\hline
\hline
$\chi^2$/dof  &  379/332 & 371/332   \\
N$_H$(10$^{22}$ cm$^{-2}$) &  2.46 [2.30 - 2.64] & 2.48 [2.32 - 2.64]  \\
H    & 1 &  1   \\
He   & 25.6  & 25.6  \\
C~   & 0.9 & 0.9   \\
N~   & 140.0 & 140.0   \\
O~   & 0.9 & 0.9   \\
Ne   & 16.9 [7.1 - 32.6] &  5.8 [0.3 - 13.7]  \\ 
Mg   & 3.5 [1.9 - 5.9]  & 1.6 [0.5 - 2.9]  \\
Si   & 5.1 [4.1 - 6.3] & 3.7 [3.1 - 4.5]  \\
S~   & 7.8 [6.5 - 9.3] & 6.4 [5.4 - 7.4]  \\
Ar   & 9.1 [5.4 - 12.9] & 8.5 [5.1 - 11.9]  \\
Ca   & 9.4 [3.4 - 15.7] & 9.0 [2.8 - 15.3]  \\
Fe   & 6.2 [4.6 - 7.9] & 6.9 [5.1 - 8.8]  \\
Ni   & 1 & 1  \\
F$^{(c)}_X$(PN)  & 1.45 (20.1) & 1.45 (28.2)  \\
F$^{(c)}_X$(MOS) & 1.56 (21.6) & 1.56 (30.4) \\
\hline
\hline
\end{tabular}
\end{center}

All abundances
are with respect to their solar values (\citealt{an_89}).
Brackets enclose 90\% confidence intervals.

$^{a}$ CSW model with decreased mass losses (0.5 of the nominal
values), stellar wind velocities scaled by a factor of 1.5 (see text).

$^{b}$ CSW model with increased binary separation (by a factor of 4),
stellar wind velocities scaled by a factor of 1.5 (see text).

$^{c}$ The observed X-ray flux (0.5 - 10 keV) followed in parentheses
by the unabsorbed value. Units are $10^{-12}$ ergs cm$^{-2}$ s$^{-1}$.

\end{table}

To explore this possibility, a set of CSW models (both in CIE and NEI) 
was considered with increased wind velocities for all the three basic 
cases ($\eta = 0.02, 0.028$~and~$0.04$). It is anticipated that
increasing wind velocities will make the NEI effects more pronounced
(see eq.~[\ref{eq:gam_nei}]).

On the other hand, 
assuming the distance to \WR is well constrained ($630\pm70$~pc,
\citealt{ch_92}) 
the mass loss or the binary separation should be changed
in order to solve the problem with the luminosity
(emission measure) mismatch.
Namely, it is necessary to decrease the
emission measure in the interaction region which can be done in two
ways: (i) by decreasing the mass loss rates; (ii) by increasing the binary 
separation. This follows from the relation between the emission
measure in the interaction region and the parameters for spherically 
symmetric stellar winds:
\begin{eqnarray}
 EM = \int{n_e n_H} dV \propto \frac{\dot{M}^2}{v_{wind}^2 D} \nonumber
\end{eqnarray}
where $n_e, n_H \propto \frac{\dot{M}}{v_{wind} D^2}$, 
and $V \propto D^3$. As we see, increasing the wind velocity also
means a smaller value for the emission measure, but we note that the
general effect of this quantity is to control the shape of the 
resultant X-ray spectrum since the wind velocities are directly related 
to the plasma temperature in the interaction region.

Thus, to discriminate between influences of various CSW parameters on
the X-ray emission from \WR, the basic cases mentioned above,
$\eta = 0.02, 0.028$~and~$0.04$, were considered in two different
ways. First, the mass loss rates of the both stellar winds were
scaled simultaneously by a factor of $0.6~(\eta = 0.02)$,
$0.5~(\eta = 0.028)$, and $0.4~(\eta = 0.04)$, while the binary separation 
had its `standard' value. Second, the mass losses kept their nominal
values while the binary separation was increased correspondingly by
a factor of $3~(\eta = 0.02)$, $4~(\eta = 0.028)$, and 
$6.25~(\eta = 0.04)$. For each of the three basic cases, a grid of
model fits was performed as the wind velocities changed with the
same scaling factor that varied between 1 and 2. Adopting such an
aprooach, namely, to scale the mass losses and/or the wind velocities
of the both winds
with the same factor has the advantage that the wind momentum ratio
remains the same. This in turn facilitates the numerical hydrodynamic
work needed prior to the spectral fits. 

Finally, it is worth noting that the decreased mass losses and the 
increased binary separation both have similar effect, namely, they
as well as the higher wind velocities  make the NEI effects 
more pronounced (the $\Gamma_{nei}$ values become smaller; 
see eq.~[\ref{eq:gam_nei}]).

Figure~\ref{fig:csw} presents the results from the fits to
the X-ray spectrum of \WR for the cases discussed above. It is
immediately seen that the CIE CSW models are not able to give
acceptable fits ($\chi^2_{\nu} > 1.45$). On the other hand, the
inclusion of NEI improves the quality of the spectral fits reaching
acceptable values of $\chi^2_{\nu} = 1.10 - 1.15$, provided the wind
velocities have values higher  than the nominal ones by a factor of 
$1.4 - 1.6$. Interestingly, if the Gorenstein (1975) conversion
formula is used, the derived X-ray absorption from the best spectral 
fits is in a good correspondence with the data for the visual 
extinction deduced from other studies. This fact is also encouraging 
for the conclusion that the developments of the CSW model presented 
in this study are in the right direction. Examples of the most
successful CSW models are shown in Fig.~\ref{fig:spec} and details
about the fits are given in Table~\ref{tab:csw}.
It is seen from the figure that the theoretical models match very well
the overall shape of the observed spectra with the exception of the
soft energy part ($E < 0.8$~keV). We believe that future data with higher
photon statistics will help resolve this issue.

Thus, the results of the analysis of the X-ray emission from \WR 
show that there are at least three requirements 
for the numerical CSW model to be successful:
(i) higher stellar wind velocities by a factor of $1.4 - 1.6$;
(ii) reduced mass losses for the stellar winds or larger binary
separation (or both); (iii) the NEI effects must be taken into account. 
We note that the needed reduction factor for the mass loss of $\sim 2$
is well within the uncertainties for this parameter of the stellar
winds in massive stars. But given the fact that the orbital parameters
(e.g., inclination angle and eccentricity) are not well constrained for 
this binary system, much smaller corrections to the mass losses might be
sufficient. 

On the other hand, the wind velocities in massive stars are
constrained much tighter usually on the basis of the P Cygni 
profiles of ultraviolet and optical spectral lines. This is not 
exactly the case of \WR. Namely, \citet{ch_92} derived a wind velocity
of 900\kms from the analysis of the HeI~$2.058 \mu$m P Cygni line
profile, and the spectral resolution was 500\kms. Also from analysis
of NIR observations with spectral resolution of 470\kms
(HeI $1.083 \mu$m emission line), \citet{ee_94} 
deduced 1100\kms for the wind velocity in \WR. Later,
\citet{ham_95} proposed a 1000\kms wind velocity, as their analysis
was based on optical HeI lines. 
Based on {\it ISO-SWS} data
with spectral resolution of 150-300\kms, a wind velocity of 950\kms was
derived from the line profiles of the forbidden lines
[NeIII] $15.5 \mu$m, [SiIV] $10.5 \mu$m, [CaIV] $3.21 \mu$m 
(\citealt{vdh_96}; \citealt{mo_00}). As we see, the wind velocity
information on \WR comes mostly from IR spectra which not always had a
very high spectral resolution. 
For a large sample of galactic WR
stars \citet{ee_94} compared the terminal wind velocities derived from 
infrared data (HeI $1.083 \mu$m and $2.058 \mu$m lines) with those from 
the conventional methods (based either on the P Cygni profiles of
resonance lines in the ultraviolet region or on the widths of the
optical lines). Interestingly, they have found that the values derived
in IR are systematically lower by $\sim 30$\%. 

If this is the case, then the wind velocity scaling factor derived here 
from the analysis of the X-ray data on the basis of the CSW model might 
not be unreasonable.  Nevertheless, it is worth noting that if no solid 
evidence for higher stellar wind velocities are found in the future, this 
will pose serious problems for the CSW scenario as a central mechanism 
for the origin of X-rays in \WR.

\section{Discussion}
\label{sec:disc}

\subsection{CSW Shocks: Collisional or Collisionless?}
All the cases considered so far in this study assume a
single-temperature hydrodynamic approximation, that is equal electron 
and ion temperatures. We recall that the thermal X-ray emission traces 
only the hot electrons, so, its analysis gives information mostly 
on the electron temperature in hot plasmas. If the plasma is heated 
through shocks, the kinetic energy of the gas flow 
is immediately (at the shock front) deposited to heavy particles 
(nucleons) and energy exchange between heavy and light
particles (electron heating) occurs downstream via Coulomb
collisions.  Thus, some time is needed for the electron and ion 
temperatures to equilibrate. This timescale depends basically 
on the electron temperature and the nucleon number density 
(\citealt{bra_58}; \citealt{sp_62}), as it is longer in plasmas with higher
temperature and lower density.

\begin{figure*}
\centering\includegraphics[width=3.0in, height=2.5in]{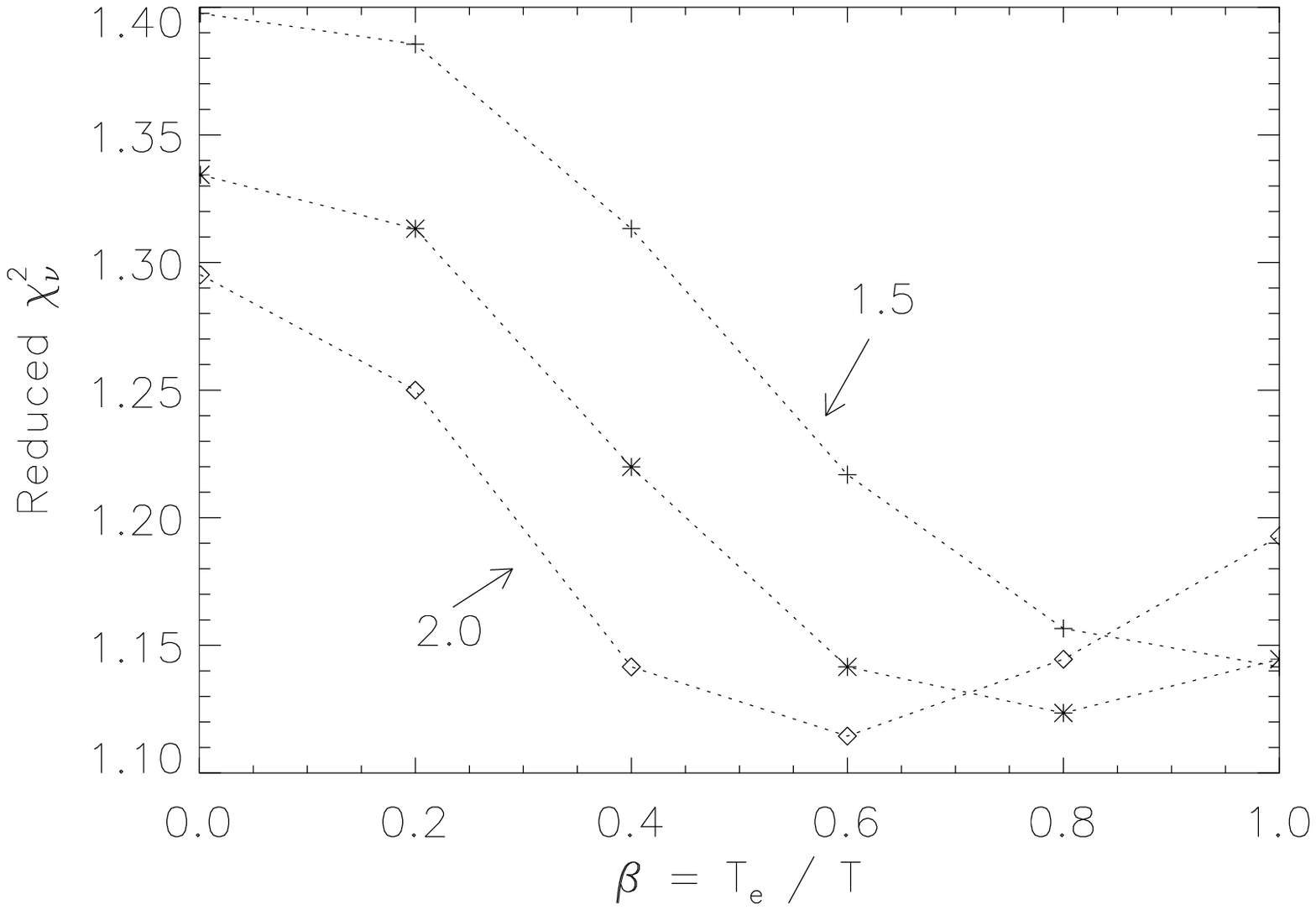}
\centering\includegraphics[width=3.0in, height=2.5in]{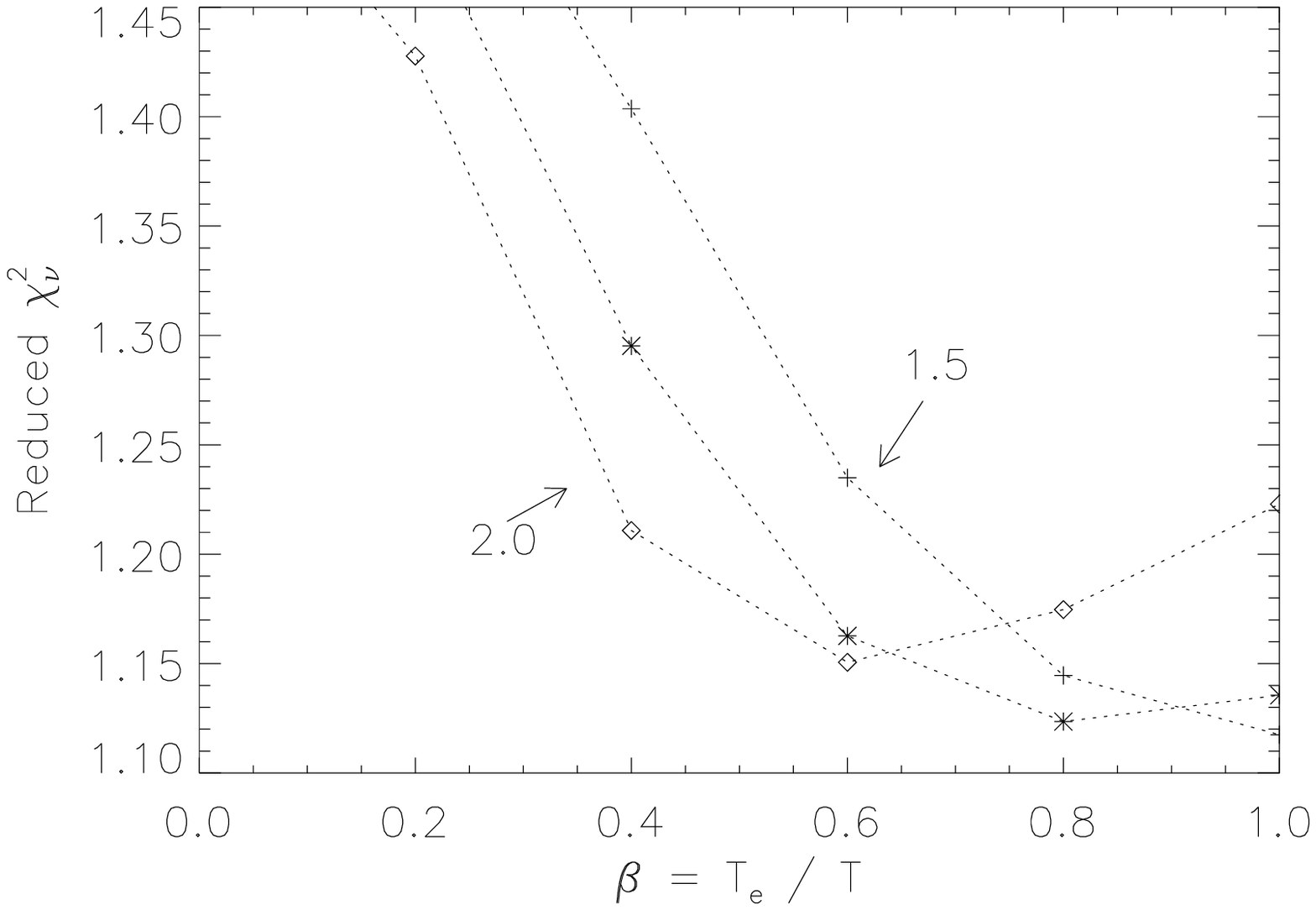}
\caption{Quality of the fit for NEI CSW models ($\eta = 0.028$) with
different degrees of initial temperature equilibration 
($\beta = T_e/T$).
Shown are cases with wind velocity increased by a factor
of 1.5, 1.7, and 2.0, respectively. 
{\bf Left panel:} Mass-loss
rates scaled while the binary separation is kept at its `nominal' value.
{\bf Right panel:} The binary separation is scaled while mass-loss rates
have their `standard' values.
The degrees of freedom in all the fits are $\nu = 332$.
}
\label{fig:beta}
\end{figure*}

Although the electron heating mechanism at the shock front is not yet
fully understood, it should be kept in mind that there are two general
cases: electron heating in {\it collisional} and {\it collisionless} shocks.
In the former case, the electrons initially attain a very low 
temperature mainly through adiabatic heating (Zeldovich \& Raizer 1967).
The situation could be very different in the latter case when various
heating mechanisms might be at work and a range of degrees of
temperature equilibration is possible up to equal electron and ion
temperatures (e.g., \citealt{mck_74}; \citealt{pap_88};
\citealt{car_88}). These possibilities are usually described by a
simple parameter that relates the electron temperature at the shock front 
to the postshock mean plasma temperature, $\beta = T_e/T$ (thus $\beta
\approx 0. - 1.$).

As the effects of unequal electron and ion temperatures are important
in high temperature, rarefied plasmas, they are thus expected to play
a role for colliding stellar winds in wide binaries. Such models
were first considered by \citet{zhsk_00} who also introduced a
dimensionless parameter that determines the relative importance of
electron heating via Coulomb collisions downstream in the interaction 
region (see eq.[1] in their work). For the nominal wind and binary
parameters of \WR (\S~\ref{subsec:wind}), the value of this
dimensioneless quantity suggests that the temperature equilibration
effects can be neglected when modelling the X-rays from this object.

On the other hand, as was shown in \S~\ref{sec:res} the
successful CSW models require higher wind velocities, and reduced mass
loss or larger binary separation than the nominal values of these
parameters. This in turn means a slower temperature equilization, that
is a higher chance for the unequal electron and ion temperatures to
affect the X-ray emission from \WR. To explore this possibility,
we ran models (both NEI and CIE) with $\beta = 0.2$, and the main
result is the lower quality of the spectral fits (see Fig.~\ref{fig:csw}).
Obviously, the reson for this is that the temperature
equalization timescale is relatively long, therefore, the electron
temperature remains below the mean plasma tempearture in the
interaction region. Guided by this and the fact that models with CIE
plasma cannot reproduce the X-ray spectrum well, more CSW models with 
different values of $\beta$ were considered as preference was given to
cases with velocity scaling factor larger than $1.5$. Some results
from the spectral fits with these NEI models are shown in 
Fig.~\ref{fig:beta}. It is seen that acceptable fits 
($\chi^2_{\nu} = 1.10 - 1.15$) are possible if the wind velocities are 
of $1.5 - 2.0$~ times larger than their nominal values, and for a high 
degree of the temperature equilization ($\beta \geq 0.4$). 
The best fits for high velocity scaling factors are obtained
with a lower initial electron temperature relative
to the mean postshock gas temperature.

\begin{figure}
\centering\includegraphics[width=3.0in, height=2.5in]{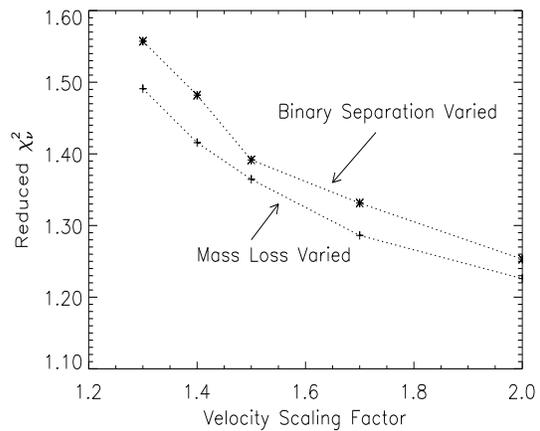}
\caption{Quality of the fit for NEI CSW models ($\eta = 0.028$) with
a very low degree of initial temperature equilibration, typical for
collisional shocks ($\beta= 0.001$).
The degrees of freedom in all the fits are $\nu = 332$.
}
\label{fig:beta_low}
\end{figure}

It is then interesting to check whether good spectral fits are possible
also for very low values of the initial electron temperature. To do so
we ran NEI CSW models with $\beta = 0.001$, a value that could be
considered typical for collisional shocks. The results are shown
in Fig.~\ref{fig:beta_low} and it is seen that no acceptable fits
are possible for the `reasonable' values of the velocity scaling
factor. 

Therefore, it is conclusive that the CSW 
shocks in \WR are {\it collisionless}!

\subsection{CSW Models vs. Discrete Temperature Models}
The analysis presented here has shown that the CSW model can explain the
{\it XMM-Newtion} X-ray data for \WR, provided that some reasonable 
adjustments to the currently accepted mass-loss parameters and orbital 
separation are made.  
On the other hand, it was shown in S07 that much simpler
models, discrete temperature ones, could be equally successful, 
as the quality of the spectral fits shows. The obvious advantage of the
CSW model is that it describes a real physical situation which is likely
present in massive binaries, but what could we learn from the 
application of these simpler models?

It is worth recalling that the interaction region in CSW binaries is
temperature stratified, thus, its X-ray emission is a sum of
individual contributions from plasmas with a range of temperatures
weighted by their emission measure. It is then not surprising that 
in such a case the total (`integrated') X-ray spectrum can be quite
well represented by isothermal plasma emission with some `average'
temperature, and its value must be lower than the maximum plasma
temperature in the shocked plasma. We note that the successful NEI
CSW models require a wind velocity scaling factor of at least $\sim 1.4$
~(see \S~\ref{sec:res}). Given the nominal wind parameters
(\S~\ref{subsec:wind}), this suggests for the shocked WR wind a maximum 
plasma temperature of $\sim 4$~keV, which is found near the line of 
centers in CSW binaries (for a strong shock with adiabatic index 
$\gamma = 5/3$ the postshock temperature is $kT = 1.956\mu v_{1000}^2$, 
keV, where $v_{1000}$ is the shock velocity in units of 1000\kms).
From such a point of view it is understandable why 
a constant temperature plane-parallel shock model with $kT \simeq 2.7$~keV
does a good job (S07). This model ({\it vpshock} in \xspec) 
has the most important 
characteristic of the CSW models for \WR, namely, it takes into account 
the NEI effects. As also discussed in S07, the fact that the
temperature of this shock model was higher than the maximum
expected for the currently accepted WR wind velocity (950\kms) was
already indicative that higher stellar wind velocities might be
present in this object. These authors reached similar conclusions on
the basis of their discrete two-temperature models with CIE.

Thus, the use of simpler models for representing the X-ray emission 
from hot plasmas can, on the one hand, be very helpful by guiding further
modelling that elaborates a real physical picture in the studied
objects. On the other hand, it may also pose some problems for
interpreting the results when models bearing quite different physics 
(e.g. CIE vs. NEI) are equally successful, as is the case for \WR.
A solid argument in favour of the NEI case here could be that the more
elaborated models are successful only if these effects are taken into
account. But, it will be much more valuable if a purely observational 
evidence is found that helps resolve this issue. For example,
future gratings data with good photon statistics will detect relatively 
strong forbidden lines in the He-like triplets of various elements if 
the NEI effects play an important role. Such observational data will
thus reveal where exactly the X-ray emission forms, i.e. in the rarified 
hot plasma of the CSW interaction region.

\subsection{CSW Models and Nonthermal Emission from \WR}
We recall that wide CSW binaries are not only sources of strong X-ray
emission but are non-thermal radio emitters as well, and \WR is no
exception to the rule. Relativistic electrons must be available in
such objects, thus, the strong shocks that shape the CSW zone in
these systems are likely places where the electrons are being accelerated. 
\citet{pitt_06} have presented this idea in a quantitative
manner developing a numerical model of the synchrotron radio emission
from CSW binaries. 
Although this model is not self-consistent it was
the first attempt to consider the non-thermal radio emission from
such objects in detail.
Naturally, their theoretical predictions were then 
confronted with the detailed radio observations of \WR. One of the
main conclusions from this analysis was that the acceleration of
relativistic electrons appear to be very efficient requiring up to
half of the thermal energy of the postshock plasma to be deposited
into non-thermal particles. We must note that if this were the case,
one of the basic consequences for the CSW zone would be a noticeable 
decrease of the temperature of the shocked hot gas. 

Contrary to this, our analysis of the {\it XMM-Newton} data for \WR 
has shown that the X-ray emitting plasma is hotter than
suggested by the nominal parameters of the stellar winds in this
object. 
It is then conclusive that there is evidence that the
CSW model is capable of explaining the X-ray emission in the wide
WR$+$OB binary \WR but further refinements are needed to successfully
describe the corresponding non-thermal radio emission from this
object.

\section{Conclusions}
\label{sec:con}
We developed an improved colliding stellar wind model that includes 
the effects of
nonequilibrium ionization in hot plasmas, and the {\it XMM-Newton}
X-ray spectra of \WR were analyzed in the framework of this model.
The basic conclusions are the following.

1. If the `standard' (\S~\ref{subsec:wind}) stellar wind parameters and 
binary separation are assumed, the CSW model predicts an emission measure
of the hot plasma that is $3-4$ times larger than what is needed
to explain X-tay emission from \WR. Also, the shape of the
theoretical spectrum does not match well that of the observed spectra,
especially, at higher energies ($E \geq 4$~keV). 
It is then concluded that hotter 
plasma must be present in the CSW zone of this object, and as
proposed by S07, such a hotter plasma could be the result of higher
wind velocities.

2. By exploring a range of stellar wind velocities, it is shown that
CSW models with nonequilibrium ionization are able to satisfactorily
fit the observed X-ray spectra of \WR, provided the stellar winds 
are faster by a factor of $1.4 - 1.6$. To resolve the issue of the 
increased emission measure, the mass loss rate should be decreased by 
a factor $\sim 2$ or the binary separation must be increased by a 
factor $\sim 4$ compared to the nominal values of these parameters. 
All these parameter changes are within their corresponding 
uncertainties.  It is worth emphasizing that in the framework of the 
CSW picture only {\bf NEI} models can correctly represent the 
observed X-ray data from this object.

3. Different electron and ion temperatures are usually found in shock 
heated rarefied plasmas, which is likely the case in wide CSW
binaries. Models that assume various degrees of temperature equalization
in the CSW shocks were confronted with the X-ray data of \WR.
Only NEI models with very high temperature equalization were
successful, a case typical for collisionless shocks. On the other
hand, models that assume a very low electron postshock temperature,
which is characteristic of collisional shocks, cannot explain the
observed data. Therefore, the analysis of the X-ray spectra of \WR
finds evidence that the CSW shocks in this object must be 
{\it collisionless}.

4. If future X-ray grating observations of \WR with good photon 
statistics are obtained, they
are expected to reveal strong forbidden lines in the He-like triplets
of various elements. The detection of strong $f$ lines would be the most 
solid argument that the hot plasma in \WR is in a state of
nonequilibrium ionization. It will then be evident that the X-ray
emission in WR 147 originates in rarefied plasma that is characteristic
of CSWs in wide binary systems.

\section{Acknowledgments}
The author would like to thank Steve Skinner for the careful reading 
of the manuscript and for his helpful comments.
This work is based on observations with {\it XMM-Newton}, an ESA
science mission with instruments and contributions directly funded by
ESA states and USA (NASA).
Also, the author appreciates the helpful comments by an
anonymous referee.

\end{document}